# Agent-based simulator of dynamic flood-people interactions

Mohammad Shirvani[1], Georges Kesserwani[1], Paul Richmond[2]

[1]*Department of Civil & Structural Engineering, University of Sheffield, Western Bank, Sheffield S10 2TN, UK*
[2]*Department of Computer Science, University of Sheffield, Western Bank, Sheffield S10 2TN, UK*

**Abstract**

This paper presents a simulator for modelling of two-way interactions between flooding and people. The simulator links a hydrodynamic agent-based model (ABM) to a pedestrian ABM in a single modelling framework on FLAMEGPU (Flexible Large Scale Agent Modelling Environment for the GPU). Dynamic coupling is achieved by simultaneous update of multiple agent types while exchanging agent information via dynamic messaging. Behavioural rules and states for the pedestrian agents are proposed related to pedestrians' presence/actions in/to floodwater, using commonly used Hazard Rate (HR) to evaluate flood risk to people and considering two roles that pedestrian can take: *evacuees*/*responders* for action *during*/*before* flooding hydrodynamics. The capability of the simulator is demonstrated over a synthetic case study of a flooded shopping centre populated by pedestrians, distinguishing two scenarios: during-flood evacuation to an emergency exit and pre-flood intervention to deploy a sandbag barrier. Our results suggest that: (1) incorporating local effects of evacuees on flood hydrodynamics can dramatically affect flood impact on the states of evacuees, and (2) the simulator can be usefully used as a tool to decide the required responders' manpower, barrier's height and risk reduction level for a safe and effective deployment strategy. Accompanying details for software accessibility are provided.

**Correspondence**
*Department of Civil & Structural Engineering, University of Sheffield, Western Bank, Sheffield S10 2TN, UK.*
Email: g.kesserwani@sheffield.ac.uk

# 1. Introduction

Flooding is a frequent hazard that can disrupt communities, in particular at urban-scale sites involving key pedestrian hubs such as leisure centres and transport infrastructures (Beker et al. 2015). Computational frameworks have become central to mitigate, prepare and manage flood risks (Lumbroso et al. 2007; Kreibich et al. 2010; Wedawatta & Ingirige 2012; Kreibich et al. 2015), and there is increasing recognition of the strategic need to particularly develop a framework for integrating human behaviour dynamics into flood risk analysis (Lumbroso & Vinet 2012; Aerts et al. 2018; Zischg 2018; McClymont et al. 2019).

Agent-Based Models (ABMs) offer a flexible platform to develop a computational framework to simulate co-evolution of the actions and interactions of multiple drivers that could lead to emergent behaviour of receptors. In recent years, ABMs have been used to support flood risk management, most commonly at meso and macro scales (Lumbroso 2008), to simulate and analyse receptors' response to floodwater. For example, ABMs were developed to evaluate risk management strategies under future climate change scenarios with multiple institutional drivers (Jenkins et al. 2017; Abebe et al. 2019), business loss and long-term effect of floods on economic growth (Li & Coates 2016; Grames et al. 2016), and the effect of protection measures, individual behaviour, and the flood occurrence frequency on the resilience of at-risk communities (Tonn & Guikema 2018). Other examples include ABMs built for flood evacuation planning in coastal areas, considering people, via vehicles, through a network of roads and with particular focus on estimating the likely number of casualties and injuries under different scenarios (Dawson et al. 2011; Mas et al. 2015; Liu & Lim 2016; Lumbroso & Davison 2018).

For flood risk analysis at urban scale, only few ABMs have been designed for evaluation of evacuation strategies considering emergent behaviour of individual people in response to a flood. Liu et al. (2009) devised an ABM to simulate evacuation of five pedestrians in underground flash floods to identify an optimal evacuation strategy by estimating the number of casualties in different scenarios. Bernardini et al. (2017) provided 'FloodPEDS', which is an ABM that incorporates a crowd of pedestrians responding to evolving flood hydrodynamics under evacuation scenarios. In FloodPEDS, pedestrian motions in floodwater is governed by adapting a standard *social force model* for evacuation dynamics (Helbing & Molnár 1995; Helbing et al. 2000) based on a (sparse) data processed from video footage of people stuck in floodwater. More generally, the Life Safety Model (LSM, www.lifesafetymodel.net) was developed to assess the risk of flood on people while taking into account dynamic interactions between multiple receptors across different scales: the LSM incorporates interactions between vehicles via a traffic model and includes a pedestrian flow model accounting for people movement as their relay on warnings to each other (Lumbroso et al. 2007; Lumbroso & Di Mauro 2008; Lumbroso et al. 2011; Lumbroso et al. 2015; Lumbroso & Davison 2018). Similarly, the LifeSIM model (www.hec.usace.army.mil/software/hec-lifesim) was developed to include pedestrians' response to emergency warnings alongside their interaction with each other and their surroundings, e.g. urban layouts and buildings, within the scope of estimating the loss of life under flood-induced evacuation conditions. However, these ABMs are suitable to only support modelling problems where the effects of people on the floodwater hydraulics do not affect the overall results (e.g. to assess the number of fatalities or injuries for violent flood types induced by tsunami or dam-break flows). In other words, they are not coupled to a hydrodynamic model treated, as ABM, to incorporate any local behavioural changes in flood hydrodynamic in response to people's move (e.g. gather) and/or action (e.g. via sandbagging).

This paper outlines the development and evaluation of a 'flood-pedestrian' simulator that dynamically couples a hydrodynamic model, treated as ABM, to a pedestrian ABM in a single computational framework. The coupled ABMs are implemented on FLAMEGPU (Flexible Large Scale Agent Modelling Environment for the GPU), which allows to define discrete and continuous agent types and to dynamically message agent-specific information between various agent types (Section 2.1). The pedestrian ABM (Section 2.3) involves continuous pedestrian agents moving on a grid of navigation agents that is aligned with the grid of flood agents involved for the hydrodynamic ABM (Section 2.2). The two-way (dynamic) coupling across the ABMs is achieved through exchange and update the information stored in the pedestrian and flood agents using the navigation agents as intermediate (Section 2.4). Behaviour rules governing pedestrian interaction with/to flood hydrodynamics are assumed and implemented based on two roles that pedestrian can take: evacuees moving in floodwater, which their presence and gathering is incorporated by systematic altering of terrain-roughness, or responders to deploy a temporary flood barrier, which take a series of actions to progressively change terrain-height via successive drop of sandbags. The dynamic coupling ability of the coupled ABMs simulator is demonstrated over a synthetic case study of a flooded and populated shopping centre considering two Scenarios: during-flood evacuation to an emergency exit (Section 3.1), and pre-flood intervention to deploy a sandbag barrier (Section 3.2). Simulation results are discussed (Section 3) considering broader implications on flood evacuation and intervention strategies for urban scale studies. Finally, conclusions are drawn (Section 4) reflecting on the future research needs to be able to validate in-model human behaviour rules to floodwater, and details of software accessibility are provided.

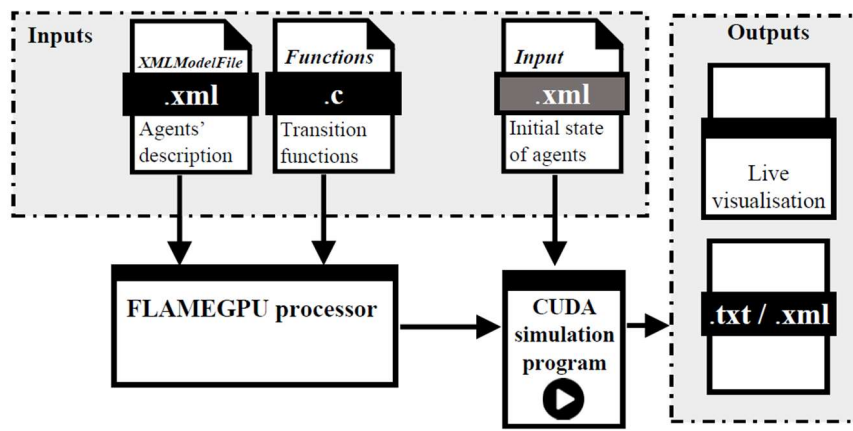

**Figure 1.** The process for generating and running an agent-based simulation program on FLAMEGPU (http://www.flamegpu.com/home).

## 2. Methodology

FLAMEGPU is a platform for the simulation of multiple agent interactions on CUDA Cores of a Graphical Processing Unit (GPU) (Richmond et al. 2009; Chimeh & Richmond 2018). It allows to create and run a CUDA simulation program by processing three inputs as shown in Figure 1. The *XMLModelFile.xml* is where a user defines formal agent specifications, including their descriptive information, type, numbers, properties, *etc.* An agent can be specified in space as either *discrete* or *continuous* (FLAMEGPU user guide). *Discrete agents* have fixed coordinates and must be pre-allocated in the memory of the GPU as 2D grid of size of a power of two number (e.g. 64 × 64, 128 × 128, 256 × 256, 512 × 512, *etc.*). *Continuous agents* change their coordinates and their population; they can be of any number as much as the GPU memory can accommodate. The *input.xml* file contains the initial conditions of the variables of state of all the defined agents. In a *single C script*, the behaviour rules to update all agents are implemented, and includes *Transition functions* to achieve dynamic message communication across multiple agents as they get simultaneously updated (FLAMEGPU user guide).

**2.1 Agents specifications**

Within a pedestrian ABM (Section 2.3), two agent types are specified: *navigation agents* and *pedestrian agents*. Navigation agents are defined to be discrete on a grid size of 128 × 128 spanning a *navigation map* on the study domain, whereas pedestrian agents are continuous so they can change positions in space and over time. Each navigation agent contains information on where pedestrian agent(s) inside it should go, namely the location of the entrances, exits and terrain obstacles, key destinations that are specified by the user as *particular* navigation agents in the *xml* files (Figure 1). Pedestrian agents receive this information, as a message from the navigation agent they are located at, as they move on the navigation map. A navigation agent can contain many pedestrian agents at a time, as they group within certain zones.

To form a hydrodynamic ABM (Section 2.2), a grid of discrete *flood agents* is defined to be of the same size as the grid of navigation agents, thus ensuring a like-for-like distribution for the flood agents and navigation agents on their respective grids. Each flood agent stores its position $x$ (m) and $y$ (m), terrain properties in terms of height $z$ (m) and roughness parameter $n_M$ (s m$^{-1/3}$), and states of floodwater variables in terms of water depth $h$ (m) and velocity components $u$ (m/s) and $v$ (m/s). The flood agent is also allocated messages to facilitate dynamic exchange of the stored information with its four neighbouring flood agents.

The coupling between the pedestrian and hydrodynamic ABMs (Section 2.4) is achieved using the grid of navigation agents as intermediate and via allocating ad-hoc messages. That is, each navigation agent is set to receive a message from the flood agent revealing its stored information. The navigation agent is set to process this information into a flood Hazard Rate (HR) quantity, and then message it to any pedestrian agent located in its area. Pedestrian agents are therefore introduced a *flood risk state* and a *walking speed state* based on the HR quantity they receive. For the pedestrian agents, they are assigned a *role* (Section 2.4) and accordingly they send a message to the navigation agent containing them at a certain time. This message informs the navigation agent on potential local amendment to the terrain properties caused by pedestrians presence or actions; namely due to local/temporal grouping of *evacuees* in certain zones leading to increasingly higher terrain-roughness, or to sandbagging by *responders* leading to progressive change in terrain-height. The navigation agent processes the messages received by the pedestrian agents, carries out an update to the terrain parameters $\{n_M, z\}$ and messages the updated information to the flood agent at its same position (Figure 3-left). Section 2.4 follows up with the details of the behaviour rules selected to govern the interactions across flood, navigation and pedestrian agents.

**2.2 Update of the grid of flood agents**

The distribution of the discrete flood agents on their grid follows similar pattern to standard mesh-based flood models using quadrilateral elements (e.g. TUFLOW-HPC, Wang et al. 2011). Hence, the numerical scheme featuring in such flood models can be integrated in the C script (Figure 1) as behaviour rules for the hydrodynamic ABM to update the states of floodwater variables stored in the flood agents. However, particular care is needed to ensure that such a scheme can operate in a *non-sequential* way to simultaneously evolve all the flood agents at a time, as required in FLAMEGPU.

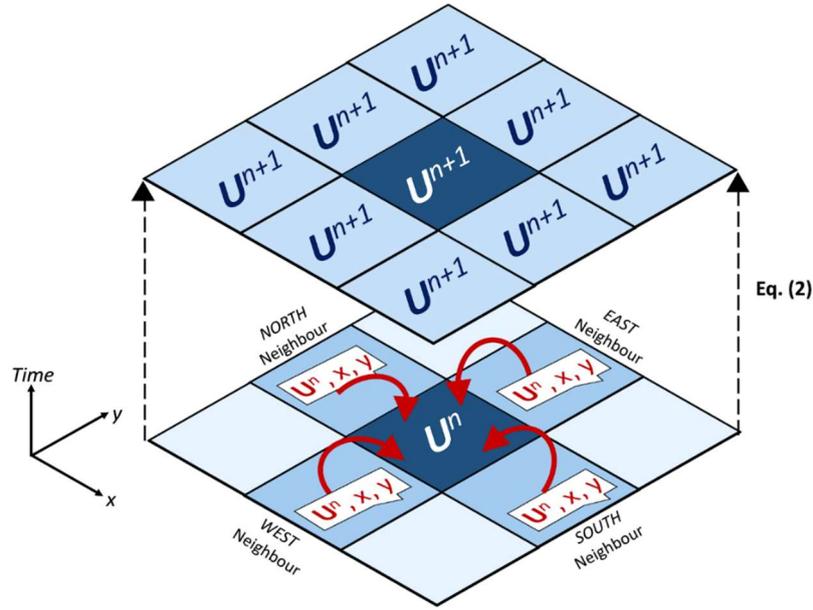

**Figure 2.** A flood agent ('dark blue') elevating its states of floodwater variables, **U**, from time iteration $n$ to $n + 1$. The process is done simultaneously for all flood agents, facilitated by the messages ('white message icons') the flood agent receives to access the states of floodwater variables of its four neighbours.

An explicit shock-capturing scheme is selected (Wang et al. 2011), and its formulation is reduced to first-order accuracy to keep the calculation stencil only dependent on the information from the immediate neighbours. The scheme numerically solves the two-dimensional (2D) depth-averaged shallow water equations, including terrain-height and -roughness as source terms, written in the following conservative form (Néelz & Pender 2009):

$$\partial_t \mathbf{U} + \partial_x \mathbf{F} + \partial_y \mathbf{G} = \mathbf{S} \qquad (1)$$

in Eq. (1), $t$ is the time coordinate, $\mathbf{U} = [h, hu, hv]^T$ is the flow vector containing the water depth and components of the unit-width flow discharge, $\mathbf{F} = [hu, hu^2 + \tfrac{1}{2}gh^2, huv]^T$ and $\mathbf{G} = [hv, huv, hv^2 + \tfrac{1}{2}gh^2]^T$ are the components of the flux vectors with $g$ being the gravity constant, and $\mathbf{S} = [0, gh(S_{0x} - S_{fx}), gh(S_{0y} - S_{fy})]^T$ is the source vector containing terrain slopes ($S_{0x} = -\partial_x z$ and $S_{0y} = -\partial_y z$) and roughness terms ($S_{fx}$ and $S_{fy}$ expressed by the Manning formula and parameter $n_M$).

For a flood agent at position $(x,y)$, their vector **U** contains constant states at time iteration $n$, which need elevating to iteration $n + 1$ according to the following formula (Figure 2):

$$\mathbf{U}^{n+1} = \mathbf{U}^n - \frac{\Delta t}{\Delta x}(\mathbf{F}_{EAST} - \mathbf{F}_{WEST}) - \frac{\Delta t}{\Delta y}(\mathbf{G}_{NORTH} - \mathbf{G}_{SOUTH}) + \mathbf{S} \qquad (2)$$

In Eq. (2), $\Delta t$, $\Delta x$ and $\Delta y$ denote the time step and dimensions of the flood agent. To elevate states of floodwater variables in the flow vector $\mathbf{U}^n$, local incoming and outgoing spatial fluxes, denoted by $\mathbf{F}_{EAST}$, $\mathbf{F}_{WEST}$, $\mathbf{G}_{NORTH}$, $\mathbf{G}_{SOUTH}$, and the source vector $\mathbf{S}$ need to be first evaluated, all of which based on reliable wetting-and-drying and terrain integration treatments (Wang et al. 2011). To simultaneously achieve these treatments and evaluations, each flood agent accesses the stored information at its neighbours, which it receives via dynamic messaging (Figure 2).

The hydrodynamic ABM on FLAMEGPU was verified in reproducing two 2D dam-break flow tests that are often used to verify hydrodynamic model implementation (Wang et al. 2011; Huang et al. 2013). In both tests, the hydrodynamic ABM (results not shown) reproduced the same predictions as the sequential counterpart, both showing very close predictions to those reported in Wang et al. (2011) and Huang et al. (2013).

Table 1. Evacuee agent states in floodwater selected based on the ranges for HR tabulated in the flood hazard matrix of the UK Environment Agency (2006).

| Hazard Rating (HR) range | | Flood risk state | Walking speed state |
|---|---|---|---|
| From | To | | |
| 0 | 0.75 | Low - safe to walk | 1.8 m/s - brisk walk |
| 0.75 | 1.5 | Medium - mildly disrupted | 0.9 m/s - slow walk |
| 1.5 | 2.5 | High - disrupted | 0.45 m/s - slower walk |
| 2.5 | 20 | Highest - trapped | 0.00 m/s - no walk |

**2.3 Update of the pedestrian and navigation agents**

The selected pedestrian ABM is already implemented on FLAMEGPU (Karmakharm et al. 2010). It adopts a standard *social force model* that is valid for modelling pedestrian flow and evacuation dynamics on dry study domains (Helbing & Molnár 1995; Helbing et al. 2000). In this pedestrian ABM, the navigation agents direct the pedestrian agents by a set of navigation rules received as messages. The walking speed of pedestrian agents in a dry area is initialised to 1.4 m/s, representative of an average human walking speed (Wirtz & Ries 1992; Mohler et al. 2007), but the existing rules allow the pedestrian agents to locally increase or decrease their walking speed to avoid collisions with each other, walls, or obstacles. The pedestrian ABM is adapted to further dynamically interact with the hydrodynamic ABM as detailed in Section 2.4.

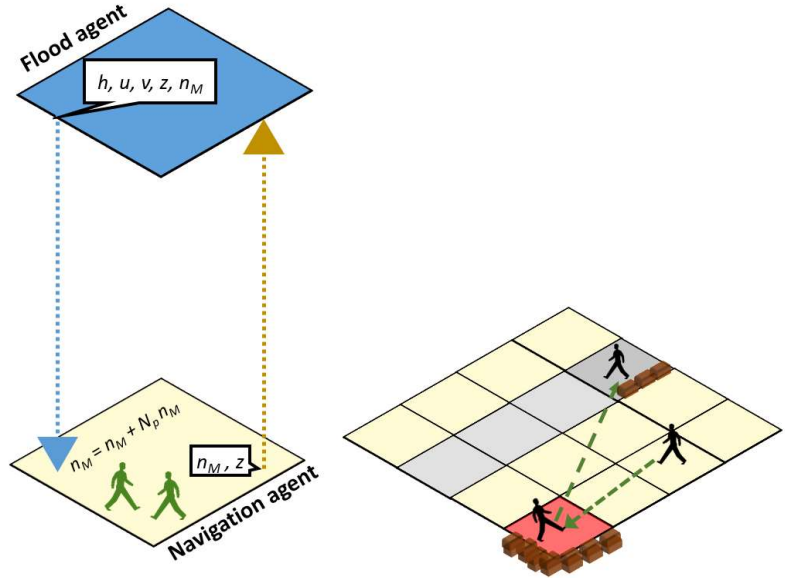

**Figure 3.** Dynamic messaging between a flood agent and pedestrian agents (evacuees) facilitated by the navigation agent located at the same position **(left)**. Messaging procedure for pedestrian agents (responders) deploying a sandbag barrier **(right)**: red navigation agent represents a 'sandbag storage' destination and grey navigation agents represent the deployment destination (Section 2.4).

**2.4 Interactions between the flood, navigation and pedestrian agents**

This section explains the behavioural rules adopted to dynamically communicate the feedback between the specified agents (Section 2.1), as they jointly update their states (Sections 2.2 and 2.3). Two different set of behavioural rules are implemented depending on the *role* assigned to the pedestrian agents (in the *xml* files, Figure 1) among either being an *evacuees* or *responders*.

*Evacuee agents* are pedestrian agents evacuating *during a flood* without a prior warning. Once a positive water depth information is received by any navigation agent on the navigation map, pedestrian agents will no longer be entering the domain, and those remaining in the domain, i.e. the evacuee agents, will be leaving to an emergency exit destination (specified by the user on the navigation map). Evacuee agents in the flooded domain receive the HR quantity from the navigation agent containing them. Estimating a flood HR usually involves measuring a product quantity of water depth $h$ and velocity magnitude $V$ (Costabile et al. 2020). Following Kvočka et al. (2016) and Willis et al. (2019), the degree of flood HR is estimated as HR = ($V$ + 0.5) × $h$, with $V$ = $max$(|$u$|, |$v$|), as also reported in the risk to people method developed for the UK Environment Agency (2006). From the HR quantity that evacuee agents receive, they are assigned a *state in floodwater* among: *low*, *medium*, *high* or *highest* based on the ranges listed in Table 1. Different *walking speed state* are also assigned for the evacuee agents, which are selected according to their *states in floodwater*, i.e. equal to: 1.8 m/s, 0.9 m/s, 0.45 m/s and 0.0 m/s, respectively (Table 1). These walking speed values are assumed[1] on the basis that: when an evacuee agent is in a low HR zone, it is able to accelerate its escape, but it will decelerate when in a medium to high HR zone and stop when located in the highest HR zone. Meanwhile, evacuee agents present in each navigation agent are counted, and their number, denoted by $N_p$, is used to update local terrain-roughness parameter by $n_M$ = $n_M$ + $N_p$ $n_M$. The updated $n_M$ is sent back to the flood agent at its location to locally and temporarily overwrite its initial $n_M$ (Figure 3-left). For this study, the initial $n_M$ value is set to be equal to 0.01, representative of clear cement (Chow 1959), and no more than 20 evacuee agents can simultaneously occupy the area of a navigation agent, which means that any local amendment in $n_M$ cannot exceed 0.2.

*Responder agents* form a group of the existing pedestrian agents, who are emergency first responders, taking a series of actions to construct a flood barrier within a specified *time window* due to an advanced flood warning. A standard sandbagging procedure is implemented to form the temporary barrier, which is an appropriate choice to support this study[2]. To further govern the moves and actions of responder agents, destinations of the sandbag storage and of the location of flood barrier are initially specified (Figure 3-right). Responder agents get information to walk to the location of the sandbag storage. Once they reach it, they are set to wait for half a minute representative of a picking up duration (specified), and to carry information on the dimension of a sandbag (Figure

---

[1] Although this assumption is sufficient to support this investigation, more sophisticated walking speed and stability rules are feasible options (e.g. Chen et al. 2019; Bernardini et al. 2020). Exploring their relative impact on pedestrian evacuation dynamics in floodwater and recovery times is currently the subject of another study.

[2] To demonstrate the feasibility of the coupled ABMs. More efficient sandbag replacement systems (Lankenau et al. 2020) can also be implemented, tested and compared in a future study.

3-right). Then they get redirected to the navigation agents spanning the temporary flood barrier, where they broadcast the sandbag dimension information and wait for half a minute representative of a safe drop out (specified). As the dimension of a sandbag is smaller than the area of a navigation agent, responder agents are set to go and share their information with one first navigation agent (specified) spanning the temporary flood barrier. The latter gather the received information until it has enough to cover one horizontal layer of sandbags all over its area, and then increments terrain-height $z$ by one unit of sandbag thickness. The process then moves to the neighbouring navigation agent spanning the flood barrier's location, and so on until the single layer of sandbags reach a wall or an existing obstacle in the domain. Responder agents then go back to the first navigation to repeat the overall process $N_L$ times until all the navigation agents spanning the flood barrier's location are filled up with $N_L$ (specified) layers of sandbags. By then, these navigation agents have already increased terrain-height by $z \times N_L$ and messaged it back to the flood agent at their location (Figure 3-left).

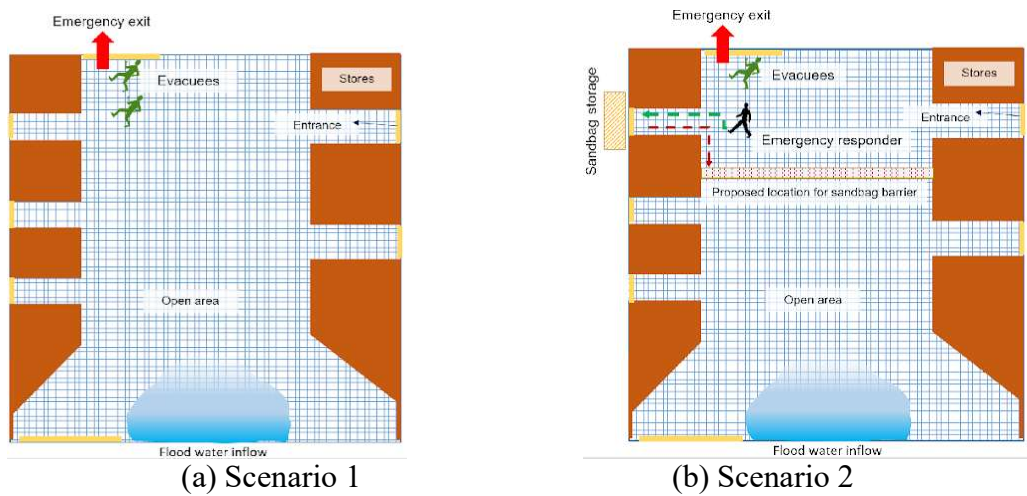

(a) Scenario 1    (b) Scenario 2

**Figure 4.** Schematic description of the hypothetical shopping centre (Section 3) with the two scenarios: (a) 'during-flood evacuation' and (b) 'pre-flood intervention'.

## 3. Demonstration on a synthetic case study

A synthetic test is proposed to evaluate the coupled ABMs for modelling dynamic interactions between people and floodwater flows. The test assumes a shopping centre encompassing a crowd of walking pedestrians that are exposed to flooding. It distinguishes two independent scenarios one involving the pedestrians as evacuees, and another involving them as responders. Scenario 1 assumes that there is no early warning nor an early evacuation plan, and studies the behaviour of pedestrians as evacuees during the propagation of the flood hydrodynamics while moving towards an emergency exit (Figure 4(a)). Scenario 2 assumes an early warning of 12 hours and studies mitigation options related to the responders' manpower and barrier's thickness for a safe and effective deployment upstream of the emergency exit (Figure 4(b)).

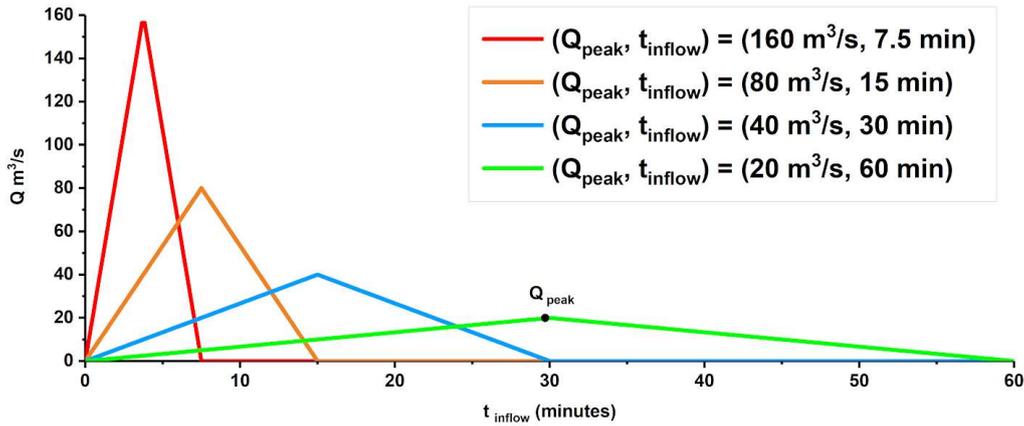

(a)

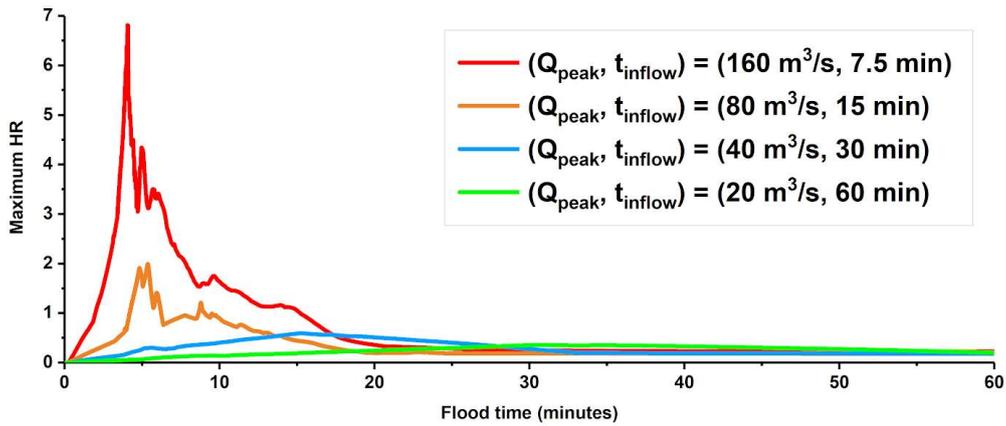

(b)

**Figure 5.** Inflow hydrographs, panel (a), and flood conditions defined by four different ($Q_{peak}$, $t_{inflow}$) pairing. The flooding conditions are defined by fixing the volume of water that will be released while proportionally doubling the discharge peak ($Q_{peak}$) and halving the duration of its occurrence ($t_{inflow}$). In panel (b), the flood conditions are analysed in term of maximum HR to assess their level of severity.

The area of the shopping centre is 332 × 332 m$^2$ (Figure 4), chosen considering the average area size of UK's 43 largest shopping centres (Gibson et al. 2018; Globaldata Consulting 2018; Sen Nag 2018; Tugba 2018). The shopping centre includes stores, located at the east and west side, separated by corridors linking the entrance doors to an open area. Through these corridors, pedestrians can enter the open area and walk toward their destinations. The open area is assumed to be occupied by pedestrians, who can enter and leave from 7 entrance doors with an equal probability of 1/7. The flood propagation occurs from the southern side (Figure 4) assuming a river inundation. As flooding starts in Scenario 1, pedestrians evacuate in response to an announcement to an emergency exit located at the northern side (Figure 4(a)) which is set to remain open during evacuation.

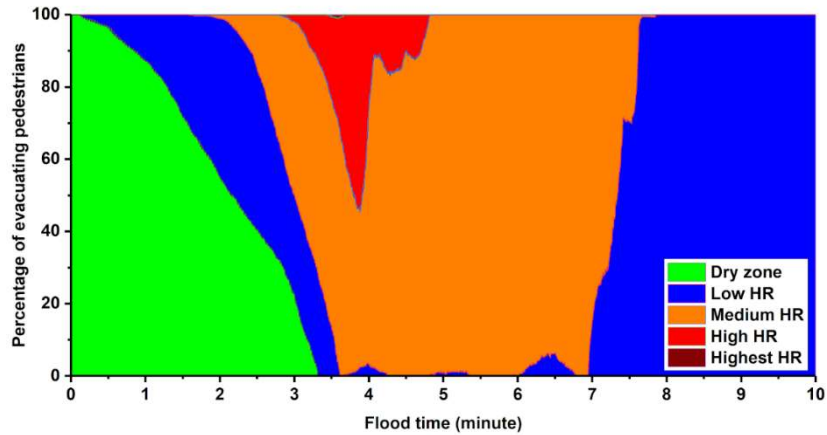

(a)

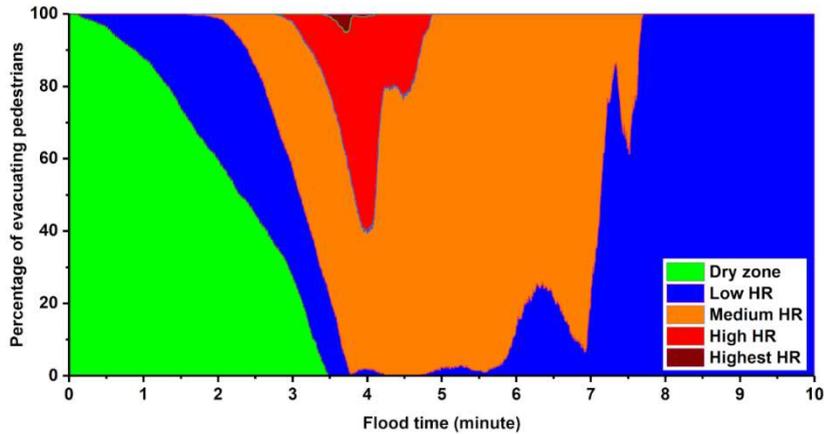

(b)

**Figure 6.** A stack chart illustrating the 'flood risk states' (Table 1) of the pedestrians as they evacuate during 10-minute flooding, without (a), and with (b) accounting for the effects of people on local flood hydrodynamics (Section 2.4).

In Scenario 2, a group of the pedestrians are responders aiming to deploy a local barrier at the location specified in Figure 4(b). The area where the barrier is intended is 168.6 m long and it has the same width as a navigation agent (i.e. 2.59 m for a grid of 128 × 128 navigation agents). The responders build the barrier by placing layers of sandbags in this area. Sandbag dimension is taken based on standard recommendations (Williamson 2010; Hellevang 2011; Padgham et al. 2014), 40 cm long × 30 cm wide × 25 cm thick. This means that 3484 sandbags are needed to form a barrier that is a one-layer thick, which is a close estimate to sandbag numbers predicted by online calculation tools (e.g. 3318 sandbags, https://sandbaggy.com/blogs/articles/sandbag-calculator), and to what is recommended in the UK official guidance (Environment Agency 2009).

In both scenarios, simulations using the coupled ABMs are run on a GPU card with 1 GB memory (Quadro K600) at a resolution of 2.59 × 2.59 $m^2$ for the grids of navigation and flood agents, and taking a population of 1000 pedestrian agents. When the domain is wet the adaptive time-step, selected under the CFL criterion (Wang et al. 2011), of the hydrodynamic ABM governed, whereas otherwise the 1.0 s time-step for pedestrian ABM is used by default.

**3.1 Flood condition selection based on HR analysis**

A flood condition is generated by an inflow hydrograph formed by pairing a peak flow $Q_{peak}$ and an inundation duration $t_{inflow}$. As the pair ($Q_{peak}$, $t_{inflow}$) is a major determining factor of flood risk, it will be considered in a triangular-shaped inflow hydrograph, which is deemed enough to support the scope of this investigation. Four flooding conditions are considered by fixing the volume of floodwater that can be released into the shopping centre and relatively selecting the pair ($Q_{peak}$, $t_{inflow}$) as shown in Figure 5(a). To ensure that the flooding conditions generated are realistic, $Q_{peak}$ for 60 min of flooding is calculated according to the initial depth and velocity reported for the Norwich inundation case study (Environment Agency 2006), which are: $h_{inflow}$ = 1 m and $v_{inflow}$ = 0.2 m/s. This corresponds to an initial pair ($Q_{peak}$, $t_{inflow}$) = (20 $m^3$/s, 60 min), with $Q_{peak} = v_{inflow} h_{inflow} B$ and $B$ = 100 m is the length of the inflow breach (see Figure 4). Three more pairs are then formed associate with increasingly more severe flood conditions, i.e. by recursive halving of $t_{inflow}$ alongside doubling of $v_{inflow}$ ($h_{inflow}$ = 1 m is fixed), which are: ($Q_{peak}$, $t_{inflow}$) = (40 $m^3$/s, 30 min), (80 $m^3$/s, 15 min) and (160 $m^3$/s, 7.5 min), respectively.

To analyse the level of severity of the selected flooding conditions, the hydrodynamic ABM is run for the four hydrograph, with slip (numerical) boundary conditions for the northern side and wall (numerical) boundary conditions for the eastern and western sides. Figure 5(b) shows the time history of the maximum HR for the four flood conditions during 60 min. With the hydrographs associated with (20 m³/s, 60 min), (40 m³/s, 30 min) and (80 m³/s, 15 min), the maximum HR does not exceeds 2 and only exceed 1 during 4 to 6 min. This indicates that these flood conditions can at worst disrupt few pedestrians for a very short duration. In contrast, the hydrograph associated with the pair (160 m³/s, 7.5 min) results in the flood condition with the highest maximum HR range, also indicating that the major flooding impact will be occurring over the first 10 min when maximum HR takes its highest values[3]. Hence, this flood condition will be considered when applying the coupled ABMs for the proposed Scenarios 1 and 2.

**3.2 Simulation of Scenario 1 (during-flood evacuation)**

The coupled ABMs is applied to simulate Scenario 1. The pedestrian ABM is set to have a constant rate of 10 entering/leaving pedestrians per entrance/exit such that to maintain a total of 1000 walking pedestrians before flooding happens. A pre-flooding duration of $t = -5$ min is set in the hydrodynamic ABM, by zeroing $Q_{peak}$, in order to allow spreading of the pedestrians all over the walkable area (blue zone in Figure 4(a)). As soon as flooding enters the domain, at $t = 0$ min, the pedestrians become evacuees and the simulation is set to terminate when all evacuees leave the domain via the emergency exit (Figure 4(a)). In a single run, the coupled ABMs simulator is set to record, every 0.1 min, the information stored in the flood agents (coordinate, water depth, water velocity and HR) and the pedestrian agents (coordinate and the HR-related flood risk states). Two runs are performed, 'with' and 'without' accounting for the effects of people on local flood hydrodynamics (Section 2.4).

---

[3] Because the aim of this study is to also explore people effects on local flood hydrodynamics, considering flood conditions that would lead to much higher maximum HR (i.e. indicative of loss of life) is out of scope.

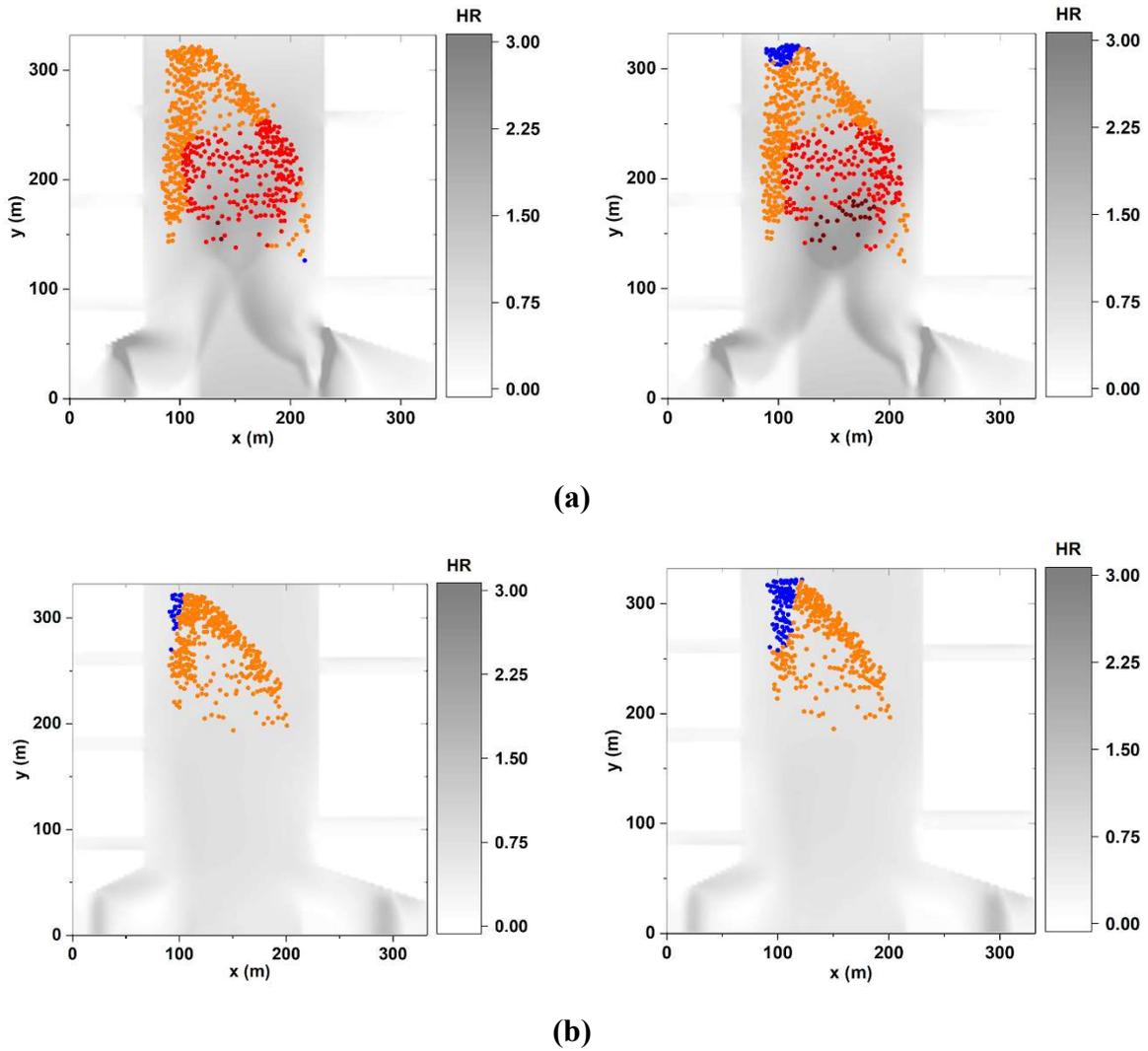

**Figure 7.** Spatial flood maps alongside the distribution of evacuees at (a) *t* = 3.6 min and (b) *t* = 6.3 min: Left and right columns contain the plots produced by the run 'without' and 'with' the effects of people on local flood hydrodynamics, respectively.

Time histories of the outputs produced by the two runs are compared in Figure 6, in terms of statistics of the flood risk states (Table 1) of evacuees. Before 2.8 min, both runs lead to almost similar statistics indicating that 60% of the evacuees are either in a dry zone (green) or in a state of low HR (blue), while the remaining 40% are at most in a medium HR state (orange). After 2.8 min and before 4.9 min, at least 55% of the evacuees have medium to highest HR states, in particular in the vicinity of 3.6 min where red to dark red areas (indicative of high to highest HR states) occupy about 5-8% wider extent for the run 'with' the effects of people on local flood hydrodynamics relative to the run 'without' these effects (compare Figure 6(a) to Figure 6(b)). After 4.9 min and before 8.0 min, the majority of the evacuees have a medium HR state as can be noted from the dominance of the orange area over this duration, namely in the vicinity of 6.3 min. Therein, a narrower orange area is observed from the run 'with' the effects of people on local flood hydrodynamics complemented by blue area that is 25% wider relative to the run 'without' these effects (compare Figure 6(a) to Figure 6(b)). After 8.0 min, all the evacuees have a low HR state (blue), irrespective of the run, and therefore are able to continue the evacuation process until it ends after 10 min. This analysis seems to suggest that: when evacuees are in a state of high to highest HR, their gathering can also increase HR states of the evacuees within their surroundings; whereas, when in a state of medium HR, gathering of evacuees can reduce the HR states of those within their surroundings.

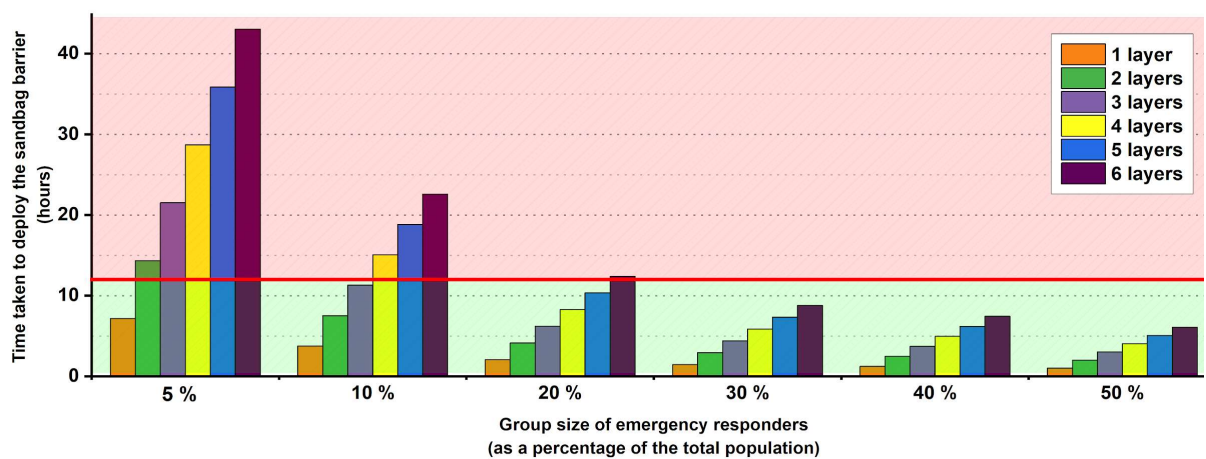

**Figure 8.** Simulated times vs. responders' group size for deploying up to six-layer thick (sandbag) barrier: 'red line' indicates flooding start time below which is safe to deploy (area shaded in 'green') or otherwise unsafe (area shaded in 'red').

This aspect is more closely explored in the spatial plots of Figure 7 for the runs 'without' and 'with' the effects of people on local flood hydrodynamics, respectively (left and right columns), after 3.6 and 6.3 min of flooding (rows (a) and (b)). The plots include the 2D spatial flood maps in terms of HR (shades of grey) and the evacuees (coloured dots representing flood risk states for the same colouring of Figure 6). Comparing the left and right columns in Figure 7(a), a clear difference is noticeable between the distribution of the evacuees and the flood maps in the crowded zones of the shopping centre: around the middle, more evacuees have high to highest HR states (red to dark red) and the local flood hydrodynamics is relatively higher (darker grey in the flood map). Whereas, closer to the emergency exit downstream, more evacuees have a low HR state (blue) indicative of relatively lower local flood hydrodynamics. The latter observation is also detected when comparing the left and right columns in Figure 7(b). Overall, these results seem to indicate that the local synergies between flood and evacuees can dramatically affect flood impact on evacuee states in floodwater.

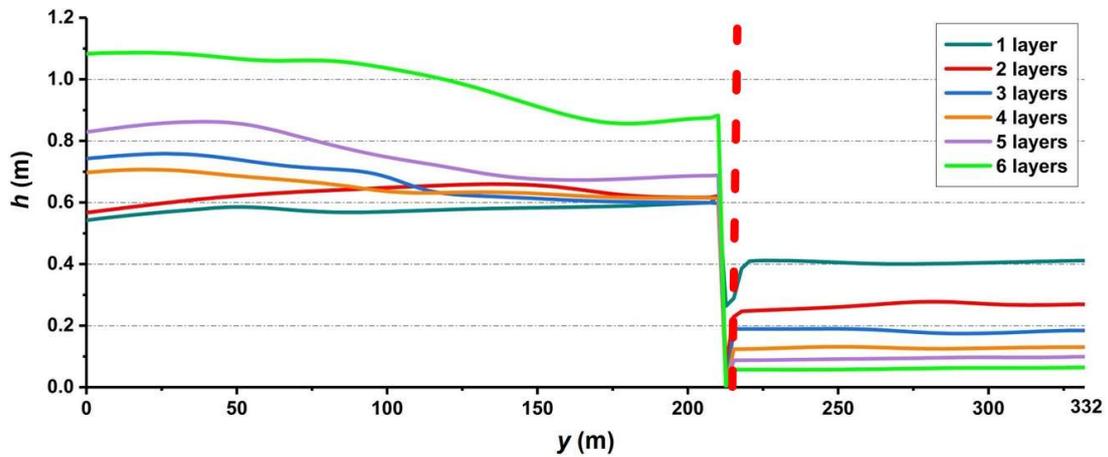

**Figure 9.** Centrelines of 2D water depth maps along *y*-axis after the deployment the sandbag barrier (red dashed line) considering up to six layers of sandbag thickness.

### 3.3 Simulation of Scenario 2 (pre-flood intervention)

The coupled ABMs simulator is re-applied for Scenario 2, with the aim to identify responders' manpower and barrier's thickness required for safe and effective deployment during 12 hrs. Six group size for the responders are explored, made of 5%, 10%, 20%, 30%, 40% and 50% of the pedestrian population, respectively, alongside six layers of thickness for the sandbag barrier. Therefore, a total of 36 simulations are run to estimate deployment time for a barrier up to six-layer thick and considering all six group sizes. Per group size, a first simulation starts with the responders evacuating as soon as they complete a one-layer thick barrier for flood risk analysis to be applied; then, by analogy, a second simulation is run to analyse the case for a two-layer thick barrier, and so on until analysing the case of a six-layer thick barrier. Analysis is also applied considering the respective changes in flood hydrodynamics in relation to the water depth and maximum HR as barrier's thickness is increased.

Figure 8 shows the simulated time taken to deploy up to a six-layer thick (sandbag) barrier for the six group sizes for the emergency responders: the time to deploy falls within the safety time window ('green' area) for a one-layer thick barrier with the 5% group, for at least three- and five-layer thick barrier with the group of 10% and 20%, respectively, and for up to a six-layer thick barrier with the 30% group and higher.[4] Figure 9 shows the changes in water depth as the barrier's thickness is increased: water depth downstream of the barrier reduces to around 0.4 m with one-layer thickness, to around 0.3 m with two-layer thickness and to less than 0.2 m with tree-layer thickness and higher. To help assess the level of safety attributed to these water depths, it is further necessary to analyse their respective velocity impacts (page 13, Environment Agency 2006).

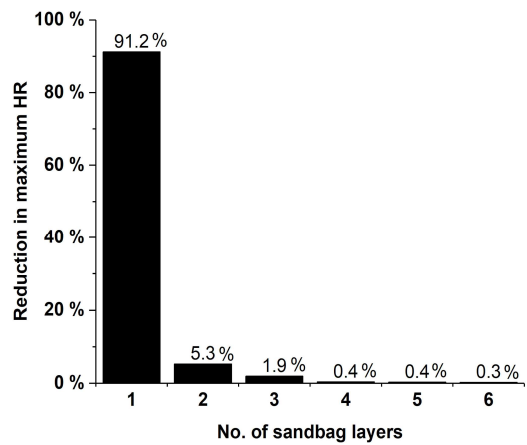

**Figure 10.** Cumulative percentage of maximum HR reduction in line with increased thickness of the barrier in terms of number of sandbag layers.

Therefore, Figure 10 illustrates the relative change in maximum HR downstream of the barrier and line with an increase in the barrier's thickness. After a one-layer thick barrier, a major drop of 91.2% in maximum HR is observed, which is quite expected relative to having no barrier at all. After two- and three-layer thickness, further relative reduction of 5.3% and 1.9%, respectively, is observed for the maximum HR. After four-layer thickness, relative reduction in maximum HR remains insignificant (close to 0.4%), suggesting that there is no need to go beyond three layers to reduce the flood risk to potentially walking pedestrians downstream of the barrier. Overall, the analyses in Figures 8-10 seem to suggest that a three-layer thick (0.75 m height) barrier would be sufficient to alleviate the flood impacts upstream of the emergency exit of the shopping centre, and robust and safe deployment within 12 hrs is feasible by at least involving the 10% group size (i.e. 100 responders).

---

[4] On a side note, no significant reduction in deployment times is observed for the three largest group sizes (Figure 8). This is likely because higher waiting times are needed in line with higher number of responders

## 4. Conclusions and outlook

The FLAMEGPU platform was used to dynamically couple validated hydrodynamic and pedestrian agent-based models (ABMs). The pedestrian ABM involved continuous pedestrian agents moving on a grid of navigation agents that is aligned with the grid of flood agents involved in the hydrodynamic ABM. The two-way coupling across the ABMs was achieved through exchange and update the information stored in the pedestrian and flood agents using the navigation agents as intermediate. Behaviour rules governing pedestrian interaction with/to flood hydrodynamics were implemented based on two roles that pedestrian can take: evacuees moving in floodwater, which their presence and gathering was incorporated by systematic altering of terrain-roughness, or responders to deploy a temporary flood barrier, which take a series of actions to progressively change terrain-height (successive drop of sandbags).

The functioning of the coupled ABMs simulator was demonstrated over a synthetic case study of a flooded and populated shopping centre considering two Scenarios: (1) during-flood evacuation to an emergency exit and (2) pre-flood intervention to deploy a sandbag barrier. Simulation results of Scenario 1 clearly identify that incorporating local effects of evacuees on flood hydrodynamics can dramatically affect flood impact on evacuees' states in floodwater. Simulation results of Scenario 2 provides evidence that the coupled ABMs can also be used to simultaneously decide the required number of manpower for emergency first responders, the required height for a temporary flood barrier, and the associated level of flood risk reduction for a safe and effective deployment strategy. These suggest a great potential for the coupled ABMs simulator for use to inform emergency flood evacuation and intervention strategies for urban scale studies.

Work is presently underway to further support and evaluate the performance of coupled ABMs simulator with experimentally-validated in-model human behaviour rules to floodwater, i.e.

variable body shapes and height for the human population, realistic people walking speeds and stability rules, and to explore its potential to plan evacuation and intervention strategies for a real study site. There is also a crucial need for interdisciplinary research across social science and psychology, hydraulic engineers and computational scientists to more realistically characterise and robustly formulate the hydro-social rules that would feature in such a flood-people simulator.

**Acknowledgments and software accessibility**

This work was supported by the UK Engineering and Physical Sciences Research Council (EPSRC) grant EP/R007349/1. The authors thank Mozhgan Kabiri Chimeh and Peter Heywood from the Research Software Engineering (https://rse.shef.ac.uk/) group for providing technical support during the implementation of the ABMs on FLAMEGPU.

The data and software that support the findings of this study are publicly available via DAFNI (www.dafni.ac.uk/) alongside a run guide document. Further description on the coupled ABMs simulation can be also found on www.seamlesswave.com/Flood_Human_ABM.

Bernardini, G., Postacchini, M., Quagliarini, E., Brocchini, M., Cianca, C., & D'Orazio, M. (2017). A preliminary combined simulation tool for the risk assessment of pedestrians' flood-induced evacuation. *Environmental modelling & software*, **96**, 14-29.

Bernardini, G., Quagliarini, E., D'Orazio, M., & Brocchini, M. (2020). Towards the simulation of flood evacuation in urban scenarios: Experiments to estimate human motion speed in floodwaters. *Safety Science*, **123**, 104563.

Chen, Q., Xia, J., Falconer, R. A., & Guo, P. (2019). Further improvement in a criterion for human stability in floodwaters. *Journal of Flood Risk Management*, **12**(3), e12486.

Chimeh, M. K., & Richmond, P. (2018). Simulating heterogeneous behaviours in complex systems on GPUs. *Simulation Modelling Practice and Theory*, **83**, 3-17.

Chow, V.T. (1959), Open channel hydraulics, McGraw-Hill, New York.

Costabile, P., Costanzo, C., De Lorenzo, G., & Macchione, F. (2020). Is local flood hazard assessment in urban areas significantly influenced by the physical complexity of the hydrodynamic inundation model?. *Journal of Hydrology*, **580**, 124231.

Dawson, R.J., Peppe, R., & Wang, M. (2011). An agent-based model for risk-based flood incident management. *Natural hazards*, **59**(1), 167-189.

Environment Agency/Defra. (2006). R&D outputs: flood risks to people: Phase 2 FD2321/TR1. The risks to people methodology. The Flood and Coastal Defence R&D Programme.

Gibson, A., Percy, J., Yates, D., & Sykes, T. (2018) UK Shopping Centres the Development Story January 2018 [online]. Cushmanwakefield. http://www.cushmanwakefield.co.uk/en-gb/research-and-insight/2018/uk-shopping-centres-the-development-story-january-2018 [accessed 10 December 2018]

Globaldata Consulting, (2018). Top 50 UK Shopping Centres Report [online]. GlobalData. https://www.globaldata.com/ground-breaking-report-sheds-new-light-uks-best-performing-shopping-centres [accessed 10 December 2018]